\documentclass[aps,prb,preprint,groupedaddress,showpacs]{revtex4}
\usepackage{amsmath}
\usepackage{graphicx}% Include figure files
\usepackage{dcolumn}% Align table columns on decimal point
\usepackage{bm}
\begin{document}
\title{Persistent current generation by small polarons in a spin-density wave background.}
\author{Hiroyasu Koizumi}
\email[ ]{koizumi@ims.tsukuba.ac.jp}
%\thanks{}
\affiliation{Institute of Materials Science, University of Tsukuba,Tsukuba, Ibaraki 305-8573, Japan}
\date{\today}

\begin{abstract}
Persistent current generation by small polarons embedded in a spin-density wave background in a two dimensional lattice is theoretically studied. When the embedded small polarons become cores of merons and antimerons (vortices in spin configuration with winding numbers $+1$ and $-1$, respectively), a spin Berry phase arises from the spiral spin texture of the merons and antimerons, and generates a spin current around the polarons.  Electric current is, however, absent due to the exact cancellation of the up- and down-spin electron currents. 
By addition of an imaginary magnetic flux that {\em negates} the fictitious magnetic flux from the spin Berry phase, the state with merons and antimerons is stabilized and non-zero electric current appears; this imaginary flux becomes a real one when side-by-side bipolaronic cores of merons and antimerons are formed, where the {\em real} flux arises spontaneously from the wave function for the motion of the system as a whole; thus,  the bipolarons serve as {\em pumps} for the persistent electric current. It is also shown that a Nernst signal stems from thermal diffusion of the bipolarons. Implications of the present results to anomalous properties of cuprate superconductors, including the enhanced Nernst signal, Drude peak, and  Fermi arc are also discussed. 
The present result suggests the appearance of the phase variable or ``Nambu-Goldstone'' mode for the superconductivity without the breakdown of the global $U(1)$ gauge invariance in the cuprates.
\end{abstract}

%03.65.Vf geometric phase
%71.10.-w   many electron theory
%72.10. -d electron transport
%72.15.Gd Glvano magnetic
%72.15.-v electron conduction
%74.20.Mn nonconventional superconductivity theory
%74.72.-h cuprate superconductors
%74.20.-z  74.20.-z Theories and models of superconducting state
%74.20.Fg BCS theory and its development
% 03.65.Ca Quantum Mechanics Formalism
% 74.20.Fg BCS theory and its developmen
%\keywords{}
\maketitle

\section{Introduction}
F. Bloch once argued that a persistent current might arise from the absence of the limit,
\begin{eqnarray}
\lim_{\Delta P \rightarrow 0} {{\Delta E} \over {\Delta P}},
\label{e1}
\end{eqnarray}
where $\Delta E$ is an increase of the total energy when a momentum in one direction $P$ is increased by $\Delta P$.\cite{Bloch}   The momentum change $\Delta P$ is brought about by multiplying a function 
$\exp(i \Delta P X \hbar^{-1})$ on the wave function $\Psi$ as
\begin{eqnarray}
\Psi'=\Psi e^{i {{\Delta P X} \over {\hbar}}},
\label{w1}
\end{eqnarray}
where $X$ denotes the coordinate of the center of gravity in the direction of $P$.
Actually, the multiplied function is a wave function for the motion of the system as a whole.

In classical mechanics, the limit corresponding to Eq.~(\ref{e1}) is given by
\begin{eqnarray}
{{\partial H} \over {\partial P}}=\dot{X}, 
\end{eqnarray}
which is zero at the minimum of the Hamiltonian $H$; thus, a  persistent current, $\dot{X}\neq0$, at the minimum is impossible.

According to Bloch, however, in quantum mechanics, the limiting process in Eq.~(\ref{e1}) may not be permissible
because of the discreteness of the momentum. Let us consider a system of $N$ electrons confined in a narrow loop of circumference $L$, and take the $x$ coordinate along the loop. The single-valuedness requirement on $\exp(i \Delta P X \hbar^{-1})$ after the increase of $x$ by $L$ demands that  $\Delta P$ must be
\begin{eqnarray}
\Delta P={{Nh} \over {L}}n,
\end{eqnarray}
where $n$ is an integer. 
If  $L$ is macroscopically large, $\Delta P$ is practically continuous, thus, a result similar to the classical one will be obtained. However, if $L$ is a microscopic length (for example, several atomic distance in a crystal), the discreteness is not negligible; then a result different form the classical one may arise.

The above argument suggests that a persistent current may arise in a system with cavity regions of  microscopic sizes. In the present work, we deal with such a system, namely, a system composed of conduction electrons and small polarons with the condition that the hopping motion of the latter is much slower than that of the former, thus, the latter provide lattice spacing size cavities for the motion of the former. Then, a macroscopic persistent current may be generated as a collection of circular currents around the small polarons.

Actually, for the persistent current generation, the existence of microscopic cavities is not enough.
A condition that stabilizes a current carrying state over a currentless state is necessary. We will show that such a condition is provided if the conduction electrons form a background spin density wave (SDW).  
Thus, the system considered in this work is the inhomogeneous system composed of the conduction electrons that form an SDW background and the small polarons that provide microscopic cavities.
The inhomogeneity considered in this work may be relevant to cuprate superconductors, where
the parent compounds of the cuprate superconductors are half-filled insulators with an antiferromagnetic SDW-like correlation; and doped holes are expected to form small polarons due to strong Coulomb interaction and strong vibronic interaction.\cite{Alexandrov95,Kamimura05,Bersuker,Wang05,Schrieffer89} 

When holes are doped in the cuprates, they are argued to become cores of merons (vortices in the spin configuration with winding number $+1$) or antimerons (vortices in the spin configuration with winding number $-1$); \cite{Gross78,John04} and the existence of such objects explains the peak splitting observed in the neutron scattering experiment very naturally. The formation of merons and antimerons around the small polarons is the key phenomenon that leads to the appearance of a persistent current, which will be shown by employing a mean field approximation that leads to an SDW state in a Hubbard model. We will show that a spin current appears when the polarons become cores of merons and antimerons due to a spin Berry phase from the spiral spin texture of the merons and antimerons.
In this respect it is worth noting that dissipationless spin current generation was recently predicted and observed in systems with spin-orbit interaction,\cite{Murakami03,Niu04,Kato04} where the spin current is produced by a spin Berry phase\cite{Berry} induced by spin-orbit interaction.

For the appearance of a persistent current, however, one more twist is necessary. Although merons and antimerons generate a spin current, electric current is absent due to the exact cancellation of the up- and down-spin electron currents. This is an example of the so-called {\em Bloch's theorem} which may be stated:
`` The ground state of a time-reversal invariant Hamiltonian is currentless''. \cite{Bohm49}
In the present work we argue that the Bloch's theorem is  circumvented through the fact that states with merons and antimerons are stabilized by an {\em imaginary} magnetic flux that ``{\em negates}'' the fictitious magnetic flux from the Berry phase; namely, if a flux that counteracts the distructive interference of the wave function around each core of meron or antimeron caused by the spin Berry phase (note that the Berry phase causes destructive interference due to $\pi$ flux through the cores), the total energy is significantly decreased. Although the introduction of such an {\em imaginary} flux is artificial, it becomes a {\em real} one when the small polarons form side-by-side bipolarons; this {\em realization of the imaginary flux} occurs because the vector potential for the imaginary flux which is written as ${\bf A}_{\rm flux}={{c \hbar} \over {2q}}\chi({\bf r})$ ($q$ is electron charge and $\chi$ is an angle-variable with periodicity $2\pi$), can be transfered from the Hamiltonian to a phase factor on the wave function written as $\exp(-i {1 \over 2} \sum_j^N \chi({\bf r}_j))$, where $N$ is the total number of electrons and ${\bf r}_j$ is the coordinate of the $j$th electron. In other words, the Bloch's theorem is circumvented by the spontaneous generation of the motion of the system as a whole given by a wave function $\exp(-i {1 \over 2} \sum_j^N \chi({\bf r}_j))$ by itself. Note that this current carrying state has a partner state that is connected by the time-reversal operation; thus, the stability of the persistent current state depends on the tunneling life time to the partner state through the energy barrier between them.

An interesting effect is predicted for the above current carrying state if the bipolarons move; such a collective mode will give rise to a time-dependence of $\chi$. Since  time variation of a phase of a wave function gives rise to a chemical potential shift,\cite{Bohm59} which for charged particles means an appearance of an electrostatic potential whose electric field is given by
${\bf E}=-{1 \over c}{ {\partial{\bf A}_{\rm flux}} \over {\partial t}}
=-{{\hbar} \over {2q}} \nabla \dot{\chi}$. 
Then, the Nernst effect is expected by the thermal diffusion of the bipolaronic cores of merons and antimerons. This may explain the anomalously large Nernst signals observed in the pseudogap phase of the cuprates.\cite{Ong05} 

As will be shown later, the present model also gives qualitative explanations for the appearance of the Drude-like peak in the optical conductivity measurement and the Fermi arc observed in the angle-resolved photoemission spectra in the cuprates; and suggests the appearance of the phase variable or ``Nambu-Goldstone'' mode for the superconductivity without the breakdown of the global $U(1)$ gauge invariance in the cuprates.

This paper is organized as follows: in the next section we discuss merons and antimerons embedded in a spin density wave background in a two dimensional square lattice using a Hubbard model. In Section III, it will be shown that a Berry phase arises from the spin rotation of conduction electrons in the spiral spin texture of merons and antimerons, and generates a spin current. In Section IV, we will add an imaginary magnetic flux that negates the fictitious magnetic flux from the spin Berry phase, and show that such a flux significantly lowers the total energy of the state with merons and antimerons. In Section V, we point out that  imaginary flux added in Section IV becomes a real one when side-by-side bipolarons are formed. In this case, the flux arises spontaneously from a wave function for the system as a whole, and circular electric currents around the bipolarons are produced. The appearance of the Drude weight is also expected in such a situation. In Section VI, we discuss the appearance of large Nernst signals by the flow of bipolarons. Lastly, in Section VII, the relevance and implications of the present results in cuprate superconductivity are discussed. The present work suggests a ``Nambu-Goldstone'' mode appears without the breakdown of the global $U(1)$ gauge invariance in the cuprates.

\section{Merons and antimerons in a spin-density wave background}
The model we are going to use is a Hubbard model with including small polaron effects of doped holes.  
The Hubbard Hamiltonian we use is given by
\begin{eqnarray}
{H}=-\sum_{\langle i,j \rangle, \sigma}
(t_{ji}c_{j\sigma}^{\dagger} c_{i \sigma}
+{\rm h.c.})
+ U\sum_{i} c_{i \uparrow}^{\dagger}c_{i \uparrow} c_{i \downarrow}^{\dagger}c_{i \downarrow},
\label{Hubbard}
\end{eqnarray}
where $c^{\dagger}_{i \sigma}$ ($c_{i \sigma}$) denotes a creation (annihilation) operator for electron with spin $\sigma$ at site $i$, and the first sum is taken over pairs of sites. 

We employ a mean field approximation that gives rise to a spin density wave (SDW) ordered electronic state with spin polarization in the $x$-$y$ plane.
Such a spin modulation is relevant for 2D organic conductors,\cite{Ishiguro} and also to the cuprates.\cite{Per-Anker05}
Electronic states are obtained from the following mean field Hamiltonian derived from Eq.~(\ref{Hubbard}):
\begin{eqnarray}
{H}_{\text mf}=&-&\sum_{\langle i,j \rangle, \sigma}( t_{ji}c_{j\sigma}^{\dagger} c_{i \sigma}+
{\rm h.c.})-U\sum_{j} \left( \Delta_j^{\ast} c_{j \uparrow}^{\dagger} c_{j \downarrow}
+ \Delta_j c_{j \downarrow}^{\dagger}c_{j \uparrow}
\right)
\nonumber
\\
&+&U\sum_j |\Delta_j|^2
+{U \over 2}\sum_j (c_{j \uparrow}^{\dagger} c_{j \uparrow}
+c_{j \downarrow}^{\dagger} c_{j \downarrow})
\label{mf0}
\end{eqnarray}
where the mean field at site $j$ is given by
\begin{eqnarray}
\Delta_j=|\Delta_j|e^{i \xi_j}= \langle c_{j \uparrow}^{\dagger}c_{j \downarrow} \rangle.
\label{mforder}
\end{eqnarray}

When the Coulomb interaction parameter $U$ is much larger than any of the absolute values of transfer integrals $|t_{ji}|$, the second sum in Eq.~(\ref{mf0}) is the dominant term.
In this situation, the representation that diagonalizes the second term is useful,
\begin{eqnarray}
-U\sum_{j} \left( \Delta_j^{\ast} c_{j \uparrow}^{\dagger} c_{j \downarrow}
+ \Delta_j c_{j \downarrow}^{\dagger}c_{j \uparrow}
\right)=-U\sum_{j} |\Delta_j|\left(  a_{j}^{\dagger} a_{j}
- b_{j}^{\dagger}b_{j}\right),
\end{eqnarray}
where  the new annihilation operators $a_j$ and $b_j$ are related to the original  by
\begin{eqnarray}
\left( \begin{array}{c}
b_{j}\\
a_{j}
\end{array}
\right)={ 1 \over {\sqrt{2}}}\left(
\begin{array}{cc}
e^{i {\xi_j \over 2}}& -e^{-i {\xi_j \over 2}} \\
e^{i {\xi_j \over 2}} & e^{-i {\xi_j \over 2}}
 \end{array}
 \right) \left( \begin{array}{c}
c_{j \uparrow}\\
c_{j \downarrow}
\end{array}
\right).
\label{transform}
\end{eqnarray}
Then, the first sum in Eq.~(\ref{mf0}) becomes
\begin{eqnarray}
-\sum_{\langle k,j \rangle, \sigma}[t_{jk}
\cos{{\xi_j-\xi_k} \over 2}(a_{j}^{\dagger} a_{k }+ b_{j}^{\dagger} b_{k})
+i t_{jk}\sin{{\xi_j-\xi_k} \over 2}(a_{j}^{\dagger} b_{k }+ b_{j}^{\dagger} a_{k})
+{\rm h.c.}].
\label{1st}
\end{eqnarray}

 At half filling in the large $U$ limit, the ground state is given by 
\begin{eqnarray}
|0\rangle=\prod_j a_j^{\dagger} |{\rm vac}\rangle.
\label{zero}
\end{eqnarray}

Now, we take into account the first nearest neighbor hopping term,
\begin{eqnarray}
-it \sum_{\langle k,j \rangle_1} \sin{{\xi_j-\xi_k} \over 2}(a_{j}^{\dagger} b_{k }+ b_{j}^{\dagger} a_{k})
+{\rm h.c.},
\label{first}
\end{eqnarray}
where a pair of first nearest neighbor sites  $k$ and $j$ is denoted as  $\langle k,j \rangle_1$.
Using the perturbation theory, we obtain the following minimal energy condition for the phase $\xi$:
\begin{eqnarray}
\sin^2{{\xi_j-\xi_k} \over 2}=1.
\end{eqnarray}
This condition is satisfied when the phase $\xi$ at site $j$ is chosen as
\begin{eqnarray}
\xi_j=\pi(j_x+j_y), 
\end{eqnarray}
where $(j_x,j_y)$ is the coordinate of the site $j$, and the lattice constant is set to unity.
The above phase yields an antiferromagnetically ordered state in accord with the well-known result at half filling. Note that the antiferromagnetic order here are not due to the Fermi surface nesting; it is due to the strong Coulomb interaction augmented by the first nearest neighbor hopping perturbation. Therefore, the mean field treatment employed here captures the physics of the Mott insulator to some extent.

Taking into account the antiferromagnetic correlation, the energy dispersion for the lower Hubbard band is calculated as
\begin{eqnarray}
e({\bf k})={ {\epsilon ({\bf k} +  {\bf Q})+\epsilon ({\bf k})}
\over 2}
-{1 \over 2}\sqrt{ \left[\epsilon ({\bf k}+ {\bf Q})-\epsilon ({\bf k})
\right]^2
+U^2},
\label{parent}
\end{eqnarray}
where ${\bf Q}=(\pi, \pi)$ and 
 the single particle
energy $\epsilon ({\bf k})$ is given by
\begin{eqnarray}
\epsilon ({\bf k})=-2t(\cos k_x + \cos k_y)-4t' \cos k_x \cos k_y+0.5U,
\end{eqnarray}
with $t$ and $t'$ being the first and second nearest neighbor transfer integrals, respectively (the first and second nearest hopping terms are included in this work).
The parameters $t$, $t'$, and $U$ are chosen to be $t'/t=-0.2$ and $U/t=8$. With these parameters the dispersion in Eq.~(\ref{parent}) resembles the one seen in a parent compound of a cuprate superconductor.
In Fig.~\ref{fig-disp}, the single particle energy dispersion $e({\bf k})$ is depicted. It shows a similar dispersion observed in the ARPES measurement for Ca$_2$CuO$_2$Cl$_2$. \cite{Ronning98}

\begin{figure}
\begin{center}
\end{center}
\caption{\label{fig-disp} Energy dispersion of single particle states for an antiferromagnetic spin density wave state $e({\bf k})$. The parameter used are $t'/t=-0.2$ and $U/t=8$ with $t=1$. The lattice constant is taken to be unity.}
\end{figure}

When holes are doped, we assume that they become small polarons due to strong electron-phonon interaction and the strong Coulomb interaction, which seems to be relevant to the cuprates.\cite{Phillips05,Wang05,Alexandrov95,Kamimura05,Bersuker,Schrieffer89} Further, we add the condition that the hopping motion of the polarons is so slow that polarons can be treated as static objects. In this situation, a part of the small polaron effects can be taken into account simply by setting the transfer integrals from and to polaron sites to zero. We will use this {\em static approximation} in the following.
Furthermore, we assume that the site energy for the polaron occupied site is such that the small polaron formation is energetically favorable; i.e., the polaron level is above the upper edge of the filled lower band. 
We achieve this by adding the term $\sum_j E_s (c_{j \uparrow}^{\dagger} c_{j \uparrow}
+c_{j \downarrow}^{\dagger} c_{j \downarrow})$ in Eq.~(\ref{mf0}), where $E_s$ is the polaron stabilization energy stems from the vibronic interaction.\cite{Alexandrov95} For simplicity, 
we take $E_s=-{1 \over 2}U$ in the following calculations; namely, we use the mean field Hamiltonian given by
\begin{eqnarray}
{H}'_{\text mf}=-\sum_{\langle i,j \rangle, \sigma}( t_{ji}c_{j\sigma}^{\dagger} c_{i \sigma}+
{\rm h.c.})-U\sum_{j} \left( \Delta_j^{\ast} c_{j \uparrow}^{\dagger} c_{j \downarrow}
+ \Delta_j c_{j \downarrow}^{\dagger}c_{j \uparrow}
\right)+U\sum_j |\Delta_j|^2.
\label{mf}
\end{eqnarray}

The electronic state is calculated using the mean-field Hamiltonian Eq.~(\ref{mf}) by self-consistently obtaining $\Delta_j$ using Eq.~(\ref{mforder}). The number of electrons is the number of sites minus the number of polarons (or holes).
In Fig.~\ref{fig-spin}(a), the spin texture for the state with the phase $\xi_j=\pi(j_x+j_y)$ is depicted; the system size is a $16\times16$ lattice, and the periodic boundary condition is employed.

\begin{figure}
\begin{center}
\end{center}
\caption{\label{fig-spin} Spin textures for states with four small polarons embedded in an antiferromagnetic background. The four small polarons are located at $(6,6), (9,6), (6,9)$, and $(9,9)$. (a): spin texture for the state without a meron or an antimeron. The total energy is $-614.5t$; (b): the same as (a) but the polarons are now cores of merons (indicated by $M$), and antimerons (indicated by $A$). The total energy is higher than that for (a) by $0.7t$.}
\end{figure}

Now, we consider the case where doped holes become cores of merons and antimerons. In general, merons and antimerons appear in meron-antimeron pairs because such paris do not severely destroy the background spin texture. Thus, the conservation of the winding number is expected since the winding number for a meron is $+1$ and that for an antimeron is $-1$.\cite{John04} 

In order to include the spin texture modulation by merons and antimerons, we write the phase $\xi$ as
\begin{eqnarray}
\xi_j=\pi(j_x+j_y)+\sum_M \tan^{-1} {{j_y-y_M} \over {j_x - x_M}}
- \sum_A \tan^{-1} {{j_y-y_A} \over {j_x - x_A}}, 
\label{meronp}
\end{eqnarray}
where $M$ and $A$ in the sums run through merons and antimerons, respectively.
In the present work the phase $\xi_j$ is fixed; we tried the optimization for $\xi_j$ by the procedure given in Ref.\onlinecite{John04}, but it did not lower the total energy in the present mean field treatment.

In Fig.~\ref{fig-spin}(b), the spin texture for the state with two merons and two antimerons is shown. The total energy for this state is $-613.8t$, while the one without a meron or an antimeron is $-614.5t$. This shows that the formation of merons and antimerons destabilizes the system. This contradicts the results in 
Ref.\onlinecite{John99} where the stabilization by the meron formation is observed; this difference probably comes from the neglect of the nearest neighbor Coulomb interaction and the inclusion of the small polaron effects in the present calculation.

\section{Spin currents generation by merons and antimerons}
In this section, we show that circular spin currents appear around polaronic cores of merons and antimerons.

Current densities are calculated using the following current density operators;
\begin{eqnarray}
j_x(l)&=&it\sum_s \left( c^{\dagger}_{l+x s}c_{l s}- c^{\dagger}_{l s}c_{l+x s}\right)
\nonumber
\\
&+&{1 \over \sqrt{2}}it'\sum_s \left( c^{\dagger}_{l+x+y s}c_{l s}- c^{\dagger}_{l s}c_{l+x+y s}
+c^{\dagger}_{l+x-y s}c_{l s}- c^{\dagger}_{l s}c_{l+x-y s}
\right),
\nonumber
\\
j_y(l)&=&it\sum_s \left( c^{\dagger}_{l+y s}c_{l s}- c^{\dagger}_{l s}c_{l+y s}\right)
\nonumber
\\
&+&{1 \over \sqrt{2}}it'\sum_s \left( c^{\dagger}_{l+x+y s}c_{l s}- c^{\dagger}_{l s}c_{l+x+y s}
+c^{\dagger}_{l-x+y s}c_{l s}- c^{\dagger}_{l s}c_{l-x+y s}
\right),
\label{currentop}
\end{eqnarray}
where $l+x$, $l+y$, $l-x$, and $l-y$ denote the nearest neighbor sites of $l$ in the $x$, $y$, $-x$, and $-y$ directions, respectively; and $l+x+y$, $l-x+y$, $l+x-y$, and $l-x-y$ denote the second nearest neighbor sites of $l$ in the $x+y$, $-x+y$, $x-y$, and $-x-y$ directions, respectively.
Note that the above operators may be deduced from the Heisenberg equation for $c^{\dagger}_{j \sigma}c_{j \sigma}$.\cite{Scalapino}
The electric current densities are obtained by the multiplication of the charge $q=-e$ on the current densities, where $-e$ is the electron charge.

The electric current densities calculated for the states with spin textures depicted in Figs.~\ref{fig-spin}(a) and (b) are zero. This is in accord with the Bloch's theorem, although the latter has nonzero spin current. In Fig.3(a), the current for up-spin electrons is depicted (the current for down-spin electrons is exactly opposite).

\begin{figure}
\begin{center}
\end{center}
\caption{\label{fig-current3} Currents for states with the spin texture depicted in Fig.~\ref{fig-spin}(b). Three different cases with and without an imaginary flux through cores of merons and antimerons are shown.  (a): current for up-spin electrons for the case without an imaginary flux. In this case the current for down-spin electrons is exactly opposite to that for up-spin electrons. The total energy is $-613.8t$; (b): the total current for the state with ${\bf A}_{\rm flux 1}$. The total energy is lower than that for (a) by $2.6t$; (c): the same as (b) but with ${\bf A}_{\rm flux 2}$. The total energy is lower than that for (a) by $3.0t$. The electric currents are obtained by the multiplication of $-e$.}
\end{figure}

The origin of the spin current can be attributed to the phase factors $e^{ i { 1 \over 2} \xi_j}$  and 
 $e^{- i { 1 \over 2} \xi_j}$ in
Eq.~(\ref{transform}), which become multi-valued when merons and antimerons are present.
A multi-valued phase factor (or a non-integrable phase factor) is a hallmark of  a ``magnetic field'',\cite{Yang74} and such a multi-valuedness can be ascribed to a {\em fictitious magnetic field}.
Let us find the vector potential for the fictitious magnetic field. We denote Wannier functions for the annihilation operator $c_{j s}$ by
$w_j$. We see that when the new operators, $a_j$ and $b_j$ in Eq.~(\ref{transform}), are employed, the phase factors $e^{ i { 1 \over 2} \xi_j}$  and $e^{- i { 1 \over 2} \xi_j}$ are multiplied on $c_{j \uparrow}$ and $c_{j \downarrow}$, respectively; this is interpreted that Wannier functions are modified as 
\begin{eqnarray}
\tilde{w}_{j \uparrow}=w_j e^{ i { 1 \over 2} \xi_j},
\end{eqnarray}
for $c_{j \uparrow}$, and
\begin{eqnarray}
\tilde{w}_{j \downarrow}&=&w_j e^{- i { 1 \over 2} \xi_j}
\end{eqnarray}
for $c_{j \downarrow}$. 
According to the prescription for the Peierls substitution,\cite{Peierls33,Luttinger}
 the above modifications of the Wannier functions are equivalent to those performed when a system is  in a magnetic field with vector potentials
\begin{eqnarray}
{\bf A}_{fic \uparrow}({\bf x}_j)=-{{c\hbar} \over {2q}} \nabla \xi_j,
\label{aup}
\end{eqnarray}
for up-spin electrons, and
\begin{eqnarray}
{\bf A}_{fic \downarrow}({\bf x}_j)={{c\hbar} \over {2q}} \nabla \xi_j,
\label{adown}
\end{eqnarray}
for down-spin electrons.
The  current for up-spin electrons is due to the fictitious vector potential ${\bf A}_{fic \uparrow}$, and that for down-spin electrons to ${\bf A}_{fic \uparrow}$; they are exactly opposite, ${\bf A}_{fic \uparrow}=-{\bf A}_{fic \downarrow}$, and produce opposite currents, ${\bf j}_{\uparrow}=-{\bf j}_{\downarrow}$. 

Strictly speaking, the transformation matrix in Eq.~(\ref{transform}) is not single-valued when
merons and antimerons exist; thus, it is not a legitimate unitary transformation.
Actually, if we follow the standard calculational procedure for the Berry phase, we obtain the same fictitious vector potentials with using a legitimate unitary transforamtion as follows:\cite{Wilczek84,Bohm03} we use
 the following single-valued transformation matrix instead of the one in Eq.~(\ref{transform})
\begin{eqnarray}
V_j = { 1 \over {\sqrt{2}}} \left( \begin{array}{cc}
e^{in \xi_j} & -e^{i(n-1) \xi_j} \\
e^{im \xi_j} & e^{i(m-1) \xi_j} 
\end{array} \right),
\label{vmj}
\end{eqnarray}
where $n$ and $m$ are integers, and $V_j$ is a single-valued unitary matrix.

Associated with the above unitary matrix is a $2 \times 2$  Berry connection
given by
\begin{eqnarray}
-iV_j \nabla V_j^{\dagger} = -{ {\nabla \xi_j} \over 2} \left( \begin{array}{cc}
2n-1& me^{i(n-m)} \\
ne^{i(m-n)} & 2m-1
\end{array} \right).
\label{vm}
\end{eqnarray}
In the following, we employ the adiabatic approximation where off-diagonal elements of the above Berry connection that causes transitions between the lower and upper bands are neglected. In the situation where the energy gap between the upper and lower Hubbard bands is much larger than external perturbations and the upper Hubbard band is empty, this approximation is a good one. Using this approximation, the motions of electrons in  the lower Hubbard band are separately treated from those in the upper one.

From Eq.~(\ref{vmj}), the annihilation operator for electrons in the lower band at site $j$ is given by
\begin{eqnarray}
a'_j ={ 1 \over {\sqrt{2}} } \left( e^{im \xi_j} c_{j \uparrow}+ e^{i(m-1)\xi_j} c_{j \downarrow} \right),
\label{eq-d}
\end{eqnarray}
and  from the $(2,2)$-element of the Berry connection in Eq.~(\ref{vm}) the fictitious vector potential for this branch is obtained  as
\begin{eqnarray}
(2m-1) {{c\hbar} \over {2q}} \nabla \xi_j.
\label{afic}
\end{eqnarray}

 For up-spin electrons, an additional vector potential $-m{{c\hbar} \over {q}} \nabla \xi_j$ arises from the phase factor $e^{im \xi_j}$ on $c_{j \uparrow}$ in Eq.~(\ref{eq-d}).
 The total vector potential for the up-spin electrons is the sum of Eq.~(\ref{afic}) and $-m{{c\hbar} \over {q}} \nabla \xi_j$, i.e., $-{{c\hbar} \over {2q}} \nabla \xi_j$, 
which agrees with Eq.~(\ref{aup}).
Likewise, for down-spin electrons, an additional vector potential $-(m-1){{c\hbar} \over {q}} \nabla \xi_j$  arises from the phase factor $ e^{i(m-1) \xi_j}$ on $c_{j \downarrow}$ in $a'_j$.
The total fictitious vector potential for down-spin electrons is given as the sum of Eq.~(\ref{afic}) and
$-(m-1){{c\hbar} \over {q}} \nabla \xi_j$, i.e.,
${{c\hbar} \over {2q}} \nabla \xi_j$
in agreement with Eq.~(\ref{adown}).

\section{Imaginary Flux states}
In the previous section, it is shown that a spin current is produced by merons and antimerons; but  the electric current is absent because the up- and down-spin currents exactly cancel each other. In this section, we add an ``{\em imaginary}'' magnetic flux through cores of merons and antimerons. If we add a suitable flux it is show that states with merons and antimerons are stabilized, and gives rise to non-zero electric current; however, we do not concern how such a flux arises in reality in this section (for this reason we call it an ``{\em imaginary}'' flux); the mechanism for the appearance of such a flux will be discussed in the next section. 

Let us apply a magnetic flux given by the vector potential
\begin{eqnarray}
{\bf A}_{\rm flux 1}({\bf x})={{c\hbar} \over {2q}} \nabla \chi_1({\bf x}),
\label{flux1}
\end{eqnarray}
where $\chi_1$ is given by
\begin{eqnarray}
\chi_1({\bf x})=\sum_M \tan^{-1} {{y-y_M} \over {x - x_M}}
-\sum_A \tan^{-1} {{y-y_A} \over {x - x_A}}.
\end{eqnarray}
The flux Eq.~(\ref{flux1}) is chosen since it may ``{\em negate}'' the fictitious magnetic field produced by merons and antimerons, and stabilizes the states with merons and antimerons. The {\em negation} here does not mean total cancelation of the effect of the fictitious magnetic field, but means the retrieval of the constructive interference of the wave functions around the cores from the destructive interference caused by the fictitious magnetic field from the Berry phase.
The negation is expected because 1)
the vector potential ${\bf A}_{\rm flux 1}$ cancels the fictitious vector potential for the up-spin electrons, 2) the sum of ${\bf A}_{\rm flux 1}$ and ${\bf A}_{fic \downarrow}$ only gives rise to an integrable $U(1)$ phase factor for the down-spin electrons.

In Fig.~\ref{fig-current3}(b), the electric current for the state with ${\bf A}_{\rm flux 1}$ is depicted. Currents are calculated using Eq.~(\ref{currentop}) but modifying the transfer integrals according to the following Peierls substitution:\cite{Peierls33,Luttinger} 
\begin{eqnarray}
t_{jk}'=t_{jk}\exp\left(iq {{ {\bf A}_{\rm flux1}({\bf r}_j)+{\bf A}_{\rm flux1}({\bf r}_k)} \over {2\hbar c}}
\cdot ({\bf r}_j-{\bf r}_k) \right).
\end{eqnarray}
Note that we will use a similar modification of the transfer integrals when we calculate current densities for states with a flux different from ${\bf A}_{\rm flux 1}$.
Significant stabilization is brought about by the introduction of the imaginary flux; the total energy is reduced by $2.6t$ for the case seen in Fig.~\ref{fig-current3}(b), which is even lower than that for the state without a merons or an antimeron by $1.9t$.

Actually, many other choices are possible for the imaginary flux that negates the fictitious magnetic field.
Next  we try the following vector potential
\begin{eqnarray}
{\bf A}_{\rm flux 2}({\bf x})={{c\hbar} \over {2q}} \nabla \chi_2({\bf x}),
\label{flux2}
\end{eqnarray}
where $\chi_2$ is given by
\begin{eqnarray}
\chi_2({\bf x})=\sum_M \tan^{-1} {{y-y_M} \over {x - x_M}}
+\sum_A \tan^{-1} {{y-y_A} \over {x - x_A}}.
\end{eqnarray}
In Fig.~\ref{fig-current3}(c), the current for the state with ${\bf A}_{\rm flux 2}$ is depicted. The current for this state is much larger than the one with ${\bf A}_{\rm flux 1}$. Its total energy is $-616.8t$, which is lower than that for the state with ${\bf A}_{\rm flux 1}$ by $0.4t$. 

In order to assess the stability brought about by the introduction of the imaginary flux, energies for three different cases with two polarons in the $16 \times 16$ lattice are calculated: i) case without a meron or an antimeron; ii) case with a meron and an antimeron but without imaginary flux; iii) case with a meron, an antimeron, and an imaginary flux. The results are listed in Table~\ref{tab:table1}. For the most of the cases, the state with a meron, an antimeron, and the vector potential ${\bf A}_{\rm flux 2}$ is the lowest in energy among the cases considered. 
Note that the energy for states without a meron or an antimeron is
almost always $-620.6t$ (or $0.3t$ above the lowest one in Table \ref{tab:table1}).
The state becomes more stable when the meron and antimeron sit closer in accord with the well-known fact  that attractive force exists between a meron and an antimeron.\cite{John04,Gross78} The most stable state is the one with an antimeron at $(7,6)$, i.e., the state with the side-by-side meron-antimeron pair, and ${\bf A}_{\rm flux 2}$. As will be discussed in the next section, a mechanism exists that gives rise to the imaginary flux in this side-by-side bipolaron case; thus, this indicates the spontaneous formation of side-by-side bipolarons with ${\bf A}_{\rm flux 2}$. It is also worth noting that between the two types of the vector potentials ${\bf A}_{\rm flux 1}$ and ${\bf A}_{\rm flux 2}$, the latter always gives slightly lower energy. 

\begin{table}
\caption{\label{tab:table1}Total energies (in the units of $t$) obtained by the mean-field Hamiltonian. The energy is relative to the lowest one, i.e., the case for $(6,7)$ with a meron, an antimeron, and ${\bf A}_{\rm flux 2}$. In each box, total energies for a case without a meron (M) or an antimeron (A), and that with a meron and an antimeron (M and A with ${\bf A}_{\rm flux 2}$) are listed. One of the two polarons is always placed at $(6,6)$ and the other at $(j_x, j_y)$ in the $16 \times 16$ square lattice. The meron core is always placed at $(6,6)$ and the antimeron core at $(j_x, j_y)$. 
For the side-by-side meron-antimeron core cases, the energy with ${\bf A}_{\rm flux 1}$ is also given. 
}
\begin{ruledtabular}
\begin{tabular}{c|c|c|c|c|c}
              &$j_x=7$&$j_x=8$&$j_x=9$&$j_x=10$&$j_x=11$\\
\hline              
$j_y=11$ & & & & &$3.0$ \\
                & & & & &$5.6 (3.5)$ \\
 \hline
 $j_y=10$& & & no M or A  &$3.0$ & $3.0$ \\
               & & &  M and A $({\bf A}_{\rm flux 2})$& $5.0 (2.6)$ & $5.3 (3.1)$ \\
 \hline           
 $j_y=9$& & no M or A &$3.0$  &$3.0$  &$3.0$  \\
              & &M and A $({\bf A}_{\rm flux 2})$ & $4.5 (1.9)$ & $4.8 (2.3)$ &$5.1 (2.7)$  \\
 \hline
 $j_y=8$& no M or A &$3.0$ &$3.0$  &$3.0$  &$3.0$ \\
             & M and A $({\bf A}_{\rm flux 1}; {\bf A}_{\rm flux 2})$ & $3.9 (1.2)$ &$4.2 (1.6)$ & $4.5 (2.0)$ 
             &$4.9 (2.3)$  \\
 \hline
 $j_y=7$& $3.0 $ &$3.0$&$3.0$&$3.0$ &$3.0$ \\
              &$3.3$ $(0.9; 0.3)$  &$3.7 (1.0)$ &$4.1(1.3)$ &$4.4 (1.8)$ & $4.8 (2.1)$ \\
 \hline
 $j_y=6$ & $2.8$& $3.0$ & $3.0$ &$3.0$ & $3.0$ \\
              & $2.9$ $(0.2; 0)$ & $3.6 (0.8)$ & $4.0 (1.1) $ &$4.4 (1.6) $ & $4.7 (2.0)$ \\
\end{tabular}
\end{ruledtabular}
\end{table}

\section{Current carrying stable states}
In this section, we show that the imaginary flux added in the previous section becomes a real one when side-by-side bipolarons are formed.

Let us take a side-by-side bipolaronic core of a meron-antimeron pair, and consider the gauge potential for
the imaginary flux for it,
\begin{eqnarray}
{\bf A}^p_{\rm flux 2}={{c\hbar} \over {2q}} \nabla \chi^p_2, 
\end{eqnarray}
where the angle variable $\chi^p$ is given by
 \begin{eqnarray}
\chi^p_2({\bf x})= \tan^{-1} {{y-y_M^P} \over {x - x_M^P}}
+ \tan^{-1} {{y-y_A^P} \over {x - x_A^P}};
\end{eqnarray}
 $(x_M^P, y_M^P)$ and $(x_A^P, y_A^P)$ denote locations of the cores of the meron and antimeron, respectively.  An example of electric current produced in the state with a side-by-side bipolaron and ${\bf A}^p_{\rm flux 2}$ is depicted in Fig.~\ref{fig-2meron}.

For every loop delineated by the hopping motion of electrons, the change of ${ 1 \over 2}\chi^p_2$ is a multiple of $2\pi$ since it encircles both of the cores, or neither of them (note that the transfer integrals from and to polaron sites are set to zero); then, the phase factor $e^{-i {1 \over 2} \sum_j^{N} \chi^p_2({\bf x}_j)}$ becomes single-valued as far as the conduction electron motion is concerend. Therefore it is legitimate to transfer the vector potential from the Hamiltonian to a wave function as
\begin{eqnarray}
\Psi'({\bf x}_1, \cdots, {\bf x}_N)=e^{-i { 1 \over 2} \sum_j^{N} \chi^p_2({\bf x}_j)}\Psi({\bf x}_1, \cdots, {\bf x}_N),
\end{eqnarray}
where the $\Psi'$ is the resulting single-valued wave function and $\Psi$ denotes the original one.
Actually, the phase factor $e^{-i { 1 \over 2} \sum_j^{N} \chi^p_2({\bf x}_j)}$ is regarded as a wave function for the motion of the system as a whole. This whole system motion will arise spontaneously since the state with it is lower in energy than that without it.

\begin{figure}
\begin{center}
\end{center}
\caption{\label{fig-2meron} Current for the state with side-by-side cores of a meron and an antimeron, and ${\bf A}_{\rm flux 2}$.
The meron core is located at $(6,6)$ and  the antimeron core is located at $(7,6)$ of the $16\times16$ square lattice in Fig.~\ref{fig-spin}. The electric current is obtained by the multiplication of $-e$.}
\end{figure}

\begin{figure}
\begin{center}
\end{center}
\caption{\label{figcurrent} Currents generated by side-by-side meron-antimeron cores, and the flux ${\bf A}_{\rm flux}$. (a): $(x_M^P, y_M^P)$=$(6,6)$, $(10,6)$,
$(7,10)$, $(12,10)$ and $(x_A^P, y_A^P)$=$(7,6)$, $(11,6)$,
$(6,10)$, $(11,10)$. The total energy is $E=-606.58t$; (b): $(x_M^P, y_M^P)$=$(6,6)$,
$(7,10)$, $(x_A^P, y_A^P)$=$(7,6)$, $(6,10)$, $(x_M^Q, y_M^Q)$=$(10,6)$, $(12,10)$,
and $(x_A^Q, y_A^Q)$=$(11,6)$, $(11,10)$. The total energy is $E=-606.08t$; (c): $(x_M^P, y_M^P)$=$(4,2)$, $(5,6)$,
$(4,10)$, $(5,14)$, $(x_A^P, y_A^P)$=$(5,2)$, $(4,6)$, $(5,10)$, $(4,14)$, $(x_M^Q, y_M^Q)$=$(12,2)$, $(13,6)$,
$(12,10)$, $(13,14)$,
and $(x_A^Q, y_A^Q)$=$(13,2)$, $(12,6)$, $(13,10)$, $(12,14)$. The electric currents are obtained by the multiplication of $-e$. }
\end{figure}

Let us consider the case where many side-by-side meron-antimeron cores exist; we adopt the gauge potential for the flux written as
\begin{eqnarray}
{\bf A}_{\rm flux}({\bf x})={{c\hbar} \over {2q}} \nabla \chi_({\bf x}),
\end{eqnarray}
where the phase $\chi$ is given by
\begin{eqnarray}
\chi({\bf x})=\sum_P \left( \tan^{-1} {{y-y_M^P} \over {x - x_M^P}}
+ \tan^{-1} {{y-y_A^P} \over {x - x_A^P}} \right)
-\sum_Q \left( \tan^{-1} {{y-y_M^Q} \over {x - x_M^Q}}
+ \tan^{-1} {{y-y_A^Q} \over {x - x_A^Q}} \right);
\end{eqnarray}
the first sum corresponds to contributions from pairs with the gauge potential analogous to ${\bf A}^p_{\rm flux 2}$ that produce clockwise currents around merons and antimerons; and the second sum to $-{\bf A}^p_{\rm flux 2}$ that produce counterclockwise currents around merons and antimerons.
In Fig.~\ref{figcurrent} examples of electric currents generated by various ${\bf A}_{\rm flux}$ are depicted.  As seen in Fig.~\ref{figcurrent}, an extended current is generated by a collection of side-by-side bipolarons.
The total energy for the state in Fig.~\ref{figcurrent}(a) is slightly lower than that for Fig.~\ref{figcurrent}(b); this seems to indicate that a larger circular current state is energetically favorable if the magnetic energy is neglected. If the magnetic energy is included a large single domain circular current state is not energetically favorable; current domains are formed to minimize the sum of the magnetic energy and total electronic energy.

Note that the current obtained here is the one outside of the conventional band theory;
the charge carriers are electrons in the fully occupied lower band.
Fig.~\ref{fig:levels} shows the plot of single particle energies for the state with the current shown in Fig.~\ref{figcurrent}(c). The lower Hubbard band is fully occupied, yet, produces the electric current.

The Drude weight is also expected in this system: the total energy is a functional of ${\bf A}_{\rm flux}$, written as $E_{\rm tot}[{\bf A}_{\rm flux}]$; the total energy depends on ${\bf A}_{\rm flux}$, and should be bounded below, thus, an optimal vector potential 
${\bf A}_{\rm flux}^{o}$ should exist; then, a general vector potential ${\bf A}_{\rm flux}$ is written as a sum of the optimized one and a deviation from it, ${\bf A}_{\rm flux}={\bf A}_{\rm flux}^{o}+\delta{\bf A}_{\rm flux}$. Assuming that the deviation $\delta{\bf A}_{\rm flux}$ is small compared with the optimized one, the total energy is written up to the second order in $\delta{\bf A}_{\rm flux}$ as
\begin{eqnarray}
E_{\rm tot}[{\bf A}_{\rm flux}] \approx E_{\rm tot}[{\bf A}_{\rm flux}^{o}]+ \int d^3 x \ d^3y 
\ ^{t}\!\left(\delta{\bf A}_{\rm flux}({\bf x})\right){{\delta^2 E_{\rm tot}[{\bf A}_{\rm flux}^{o}]} \over {\delta {\bf A}_{\rm flux}({\bf x})
\delta {\bf A}_{\rm flux}({\bf y})}} \delta{\bf A}_{\rm flux}({\bf y}).
\label{Drudef}
\end{eqnarray}
Thus, the Drude weight (in the $x$ direction) is calculated as\cite{Kohn}
\begin{eqnarray}
\pi c^2  {{d^2 E_{\rm tot}[{\bf A}_{\rm flux}^{o} + A_x {\bf e}_x]} \over {d A_x^2}} {\Big |}_{A_x=0}
=
\pi c^2 \int d^3 x \ d^3y 
\ {{\delta^2 E_{\rm tot}[{\bf A}_{\rm flux}^{o}]} \over {\delta ({A}_{{\rm flux} })_x({\bf x})
\delta ({A}_{{\rm flux} })_x({\bf y})}},
\label{Drude}
\end{eqnarray}
where ${\bf e}_x$ is the unit vector in the $x$ direction.
The Drude weight describes the sensitivity of the system with twisting the periodic boundary condition; 
if a collection of circular currents around the bipolarons  extends over the sample, the nonzero Drude weight is expected.

\begin{figure}
\begin{center}
\end{center}
\caption{\label{fig:levels} Single particle energies for the state with the current shown in Fig.~\ref{figcurrent}(c). The unit of energy is $t$. The levels in the lower Hubbard band are fully occupied, and those in the upper Hubbard band and in-gap states are empty. The in-gap states are due to small polarons whose position depends on the polaron stabilization energy $E_s$; in the present work it is chosen as $E_s=-{1 \over 2}U$.}
\end{figure}

In the current carrying state considered in this section, the Bloch's theorem is circumvented by the spontaneous generation of the whole system motion by itself; it produces a vector potential that negates the fictitious magnetic field from the spin Berry phase, and stabilizes the state with bipolaronic cores of merons and antimerons. Since the Hamiltonian is time-reversal invaraint, this current carrying state has a partner state that is connected by the time-reversal operation. These two states may be viewed as states at two minima of a double-well potential. A current carrying state is stable if the energy barrier between them is so large that the tunneling motion between them is practically negligible.

Lastly in this section, we briefly discuss the response of the system to an external magnetic field. The state with ${\bf A}_{\rm flux}$ and that with $-{\bf A}_{\rm flux}$ have the same energy and produce mutually opposite currents; when an magnetic field is applied in the $z$ direction, their energies become different. 
Let us apply a magnetic flux density $B{\bf e}_z$ in the $z$ direction to the state (${\bf e}_z$ is the unit vector in the $z$ direction) with the current shown in Fig.~\ref{figcurrent}(a); the result is  the total energy becomes $-606.55t$ for $B=-0.001$, and $-606.47t$ for $B=0.001$ where $B$ is given in the units of $2\Phi_0/a^2$ with $a$ being the lattice constant and $2\Phi_0$ equal to $hc/e$=$2.07\times 10^{-7}$ gauss$\cdot$cm$^{2}$. This result shows that the one with the diamagnetic current to the external magnetic field is more stable (the energy difference $0.08t$ is about $200$ K for the cuprates). In many cases (but not always) we have examined, the state with a diamagnetic current seems to become slightly lower than that with a paramagnetic current; this may imply an appearance of a large diamagnetic response.

\section{Nernst signals by the flow of merons and antimerons}
When the flow of bipolaronic cores of merons and antimerons occurs, the phase $\chi$ becomes time-dependent.
This means that conduction electrons are under the influence of an electric field given by
\begin{eqnarray}
{\bf E}=-{1 \over c}{ {\partial{\bf A}_{\rm flux}} \over {\partial t}}=-{{\hbar} \over {2q}} \nabla \dot{\chi}.
\label{efield}
\end{eqnarray}
In this section we show that this electric field gives rise to Nernst signals.

To calculate the electric field produced by a thermal gradient $-\nabla T$, let us take a rectangular system shown in Fig.~\ref{fig:nernst} with the length $L_x$ in the $x$ direction $(0 \leq x \leq  L_x)$, and the width $L_y$ in the $y$ direction $(0 \leq y \leq  L_y)$; a magnetic field is applied in the $z$ direction; and the thermal gradient is in the $x$ direction $-\partial_x T$. 

From Eq.~(\ref{efield}), the voltage across the width $L_y$ measured at $x={{L_x} \over 2}$ is calculated as 
\begin{eqnarray}
V_y &=&E_yL_y
={{\hbar} \over {2e}} \int_0^{L_y} dy {{\partial} \over {\partial y}}\dot{\chi}\left({{L_x} \over 2},y\right) 
\nonumber
\\
&=&{{\hbar} \over {2e}}\left[\dot{\chi}\left({{L_x} \over 2}, L_y\right)\!-\!\dot{\chi}\left({{L_x} \over 2},0\right) \right].
\end{eqnarray}

After the movement of a bipolaron from $x=0$ to $x=L_x$, the difference of the phase $\chi$ across the width at $x={{L_x} \over 2}$ changes. If $L_x$ is sufficiently large, the phase change is $-2\pi$ if $\chi$ is the one produces a clockwise circular current ($2\pi$ for the one produces a counterclockwise circular current) around a meron or an antimeron. In the situation where the bipolaron dissociation and recombination processes are frequent, the above phase change should be regarded as an average over many disconnected flows of bipolarons. Similarly, in an average sense, we denote the velocity of the flow by $v$, and use $L_x/v$ as the average time duration for the movement of a bipolaron from $x=0$ to $x=L_x$.

\begin{figure}
\begin{center}
\end{center}
\caption{\label{fig:nernst} Experimental set-up for the Nernst signal measurement.}
\end{figure}

Then, the voltage measured in the Nernst experiment is calculated as
\begin{eqnarray}
V_y
={{\hbar} \over {2e}}{{2\pi N_m} \over {L_x/v}},
\end{eqnarray}
where $N_m$ is the difference of the numbers for diamagnetic circular current cores ($N_d$) and paramagnetic circular current cores ($N_p$) of polarons, $N_m=N_d-N_p$. $N_m$ is positive in the presence of the applied magnetic field because diamagnetic current states are expected to be more stable than paramagnetic current states.

Finally, the electric field produced is calculated as
\begin{eqnarray}
E_y={{V_y} \over {L_x}}={{hv} \over {2e}}{{N_m} \over {L_x L_y}}={{hvn_m} \over {2e}},
\label{eq:nernst}
\end{eqnarray}
where $n_m$ is the surface density of $N_m$ given by $N_m/L_xL_y$. Since a loop current gives rise to a magnetic moment, circular currents around the bipolarons produce magnetization perpendicular to the two dimensional plane. If we assume that each meron or antimeron core in a bipolaron contributes to roughly the same magnetic moment, Eq.~(\ref{eq:nernst}) indicates that the Nernst signal is roughly proportional to the magnetization. Actually, the Nernst signal proportional to the magnetization has been observed in the pseudogap phase of the cuprate superconductors.\cite{Ong05}

\section{Discussion}
In this section, we shall discuss the relevance and implications of the present work in the cuprate superconductivity.

We have been dealing with the inhomogeneous system composed of conduction electrons and small polarons; the former constitute a background SDW order, and the latter become embedded bipolaronic cores of merons and antimerons. For lightly doped cuprates, the strong evidence of polaron formation exists;\cite{Gunnarsson05,Shen05} and the evidence of polaronic charge carriers also exits.\cite{Goodenough95,Alexandrov95} 
 
 The single particle levels depicted 
in Fig.~\ref{fig:levels} show the splitting of the original band into the lower and upper Hubbard bands, and between them are the in-gap states due to polarons.
 In the lightly doped cuprates, the constant chemical potential with changing the hole concentration has been observed; it is interpreted due to the presence of in-gap states that pin the Fermi level.\cite{Sawatzky94,Harima03} Spectral weight transfer that appears to be between the in-gap states and the lower Hubbard band has been observed. \cite{Ino00} 
Quite often, the charged stripes are argued to be responsible for the chemical potential pinning;\cite{Ino02} and the splitting of peaks in the neutron scattering experiments has been taken as the evidence for the existence of them. \cite{Yamada98,Tranquada}  In contrast to it the present work suggests that the chemical potential pinning is atributed to the small polaron formation; those polarons form merons and antimerons around them and explain the neutron scattering peak splitting;\cite{John04} thus, what has been considered to be due to charged stripes may be actually due to merons and antimerons whose cores are small polarons. 

It has been though that a disconnected Fermi surface like feature (called ``Fermi arc'')\cite{Marshall96,Yoshida03} seen in the angle resolved photoemission spectroscopy (ARPES) 
and  the Drude-like peak observed in the optical conductivity measurement\cite{Uchida91,Basov05} indicate that the charge carriers are not small polarons, thus, small polarons are thought to be not relevant to cupreate superconductivity. However, the present model explains those phenomena with the small polaron as an essential ingredient, although it is not the charge carrier.

Let us first simulate the ARPES experiment: we calculate the ARPES plot  as a sum of the square of the absolute value of the Fourier transform of single particle level wave functions given by
\begin{eqnarray}
I({\bf k})=\sum_{i=i_1}^{i_2} |\langle {\bf k}| \phi_i \rangle|^2,
\label{arpes}
\end{eqnarray}
where $i_1$ is the arbitrarily chosen lower limit of the occupied state, $i_2$ is the highest occupied state in the lower Hubbard band, and $\phi_i$ is the $i$th single particle wave function. The results are depicted in Fig.~\ref{fig:arpes}.
A Fermi arc which is similar to the one observed in ARPES\cite{Yoshida03} is seen in Fig.~\ref{fig:arpes}. 
The Fermi arc is more arc-like when the circular currents are all in the same (clockwise or counterclockwise) direction and produce a large current region. Therefore, the experimentally obtained Fermi arc suggests the presence of bipolaronic cores of merons and antimerons with a large circular current area, which is in accord with the result in Section IV that a larger current state is energetically favorable.

Next, we shall consider the Drude peak problem. The Drude weight of Eq.~(\ref{Drude}) is expected to arise in states with circular currents around small polarons. Low energy excitations associated with flipping of circular currents (from clockwise one to counterclockwise one, or vise versa) exist; an estimate from the case depicted in Fig.~\ref{fig:arpes} is that it is of order of $0.2t \approx 40$ meV. Therefore, the Drude-like peak observed in the optical conductivity measurement\cite{Uchida91,Basov05} is qualitatively explained in the present model.

\begin{figure}
\begin{center}
\end{center}
\caption{\label{fig:arpes} Simulations of the angle-resolved photoemission spectra (ARPES) for cuprates. Each row has three plots; Spin texture plot, the total current plot, and ARPES simulation plot (symmetrized about the $(0,0)-(\pi,\pi)$ reflection line), from left to right.
The first and second rows are those for states with one bipolaronic core. The ARPES plot  is calculated using Eq.~(\ref{arpes}) with the 15 highest occupied single particle level wave functions (energy range is from $-3.92t$ to $-3.80t$ for the first row and $-3.93t$ to $-3.80t$ for the second).
The total energy for the first row case is lower than that for the second row case by $0.2t$.
The third row is for the state with 16 bipolaronic cores. In this case, the Coulomb interaction parameter is reduced from $U=8t$ to $U=4t$; this change causes a slight elongation of the arc compared with the $U=8t$ case. The ARPES plot is obtained using Eq.~(\ref{arpes}) with the 25 highest occupied single particle level wave functions (energy range is from $-1.90t$ to $-1.61t$).}
\end{figure}

In the cuprates, the anomalously large Nernst signal and enhanced diamagnetism have been observed at temperatures well above $T_c$. \cite{Ong01,Ong05} A large Nernst signal is usually associated with the flow of Abrikosov vortices, thus, it is tempting to regard the enhanced Nernst signal as evidence for the spontaneous formation of moving Abrikosov vortices. Besides, it is also in accord with the scenario that the onset of superconductivity corresponds to the establishment of a long-range coherence of the order parameter phase.\cite{Emery95} However, it has been argued that the presence of a supercurrent that is capable of forming Abrikosov vortices seems to be implausible at such high temperatures since the sharp resistive and magnetic transitions at $T_c$ cannot be explained by the fluctuation scenario.\cite{Alexandrov04} Indeed, the scanning SQUID microscopy observed the disappearance of the Abrikosov vortices above $T_c$.\cite{Iguchi01}
The origin of the large Nernst effect is a very controversial issue.\cite{Capan05}
The present work provides a new explanation: it is due to the flow of the bipolaronic cores of merons and antimerons. The comparison of the experimental results\cite{Ong05} and the formula in Eq.~(\ref{eq:nernst}) indicates that the magnitude of the magnetization increases exponentially with the increase of $1/T$, which may be attributed to the increase of the number of the bipolarons; if we assume each bipolaron contributes either an average up or down magnetic moment produced by the circular current around it,
the magnetization will be proportional to the difference of the numbers of the up and down magnetic moments per volume.
Let us try a fit to the experimentally measured magnetization to the function $M(T)=-Ae^{-T_f/T}$, where $A$ and $T_f$ are fitting parameters; the fit is very good as shown in Fig.~\ref{fig:ong}. From the fit, the formation energy $-T_f$ of the bipolaron is estimated to be 75 meV, which is close to the value where a kink is observed in the energy dispersion of angle resolved photoemission spectrum.\cite{Lanzara} 
Note that this energy seems to be larger than that is expected for the flipping of the circular currents; and may correspond to the formation energy of a bipolaron from two separate polarons.

\begin{figure}
\begin{center}
%s\includegraphics[scale=0.5]{9x016.ps}
\end{center}
\caption{\label{fig:ong} Temperature dependence of the magnetization for the $x=0.16$ (Bi 2212) sample obtained by Y. Wang et al. \cite{Ong05} Filled circles are experimental data.
The fitting function is $M(T)=-Ae^{-T_f/T}$ with $A=0.00033$ A/m and $T_f=-1,100$ K.}
\end{figure}

In terms of the present model, the experimental fact that the observed Nernst signal above $T_c$ smoothly connects to that by the Abrikosov vortices below $T_c$\cite{Ong05} implies a rather surprising conclusion: the flow of the bipolarons smoothly connects to the Abrikosov vortex flow, thus, the phase that generates the supercurrent below $T_c$ (the phase of the order parameter) is actually the phase that generates a circular current around the bipolaron above $T_c$ (the phase for the gauge potential ${\bf A}_{\rm flux}$). In other words, the cuprate superconductivity may be the superconductivity with the bipolarons serving not as the charge carriers, but as {\em pumps} for the supercurrent.

One may think that in such a theory, the flux quantization in the units of ${{ch} \over {2e}}$,
or the ac Josephson effect with the frequency $2eV/h$ ($V$ is the applied voltage) will not be possible
because the charge on the current carrier is not $2e$. However, as will be explained below, the present theory will explain the observed flux quantization and Josephson frequency. Note that
$2e$ in ${{ch} \over {2e}}$ or $2eV/h$ does not exclusively mean the existence of $2e$ charge carriers.\cite{Koizumi99}

Let us examine the flux quantization problem: we first show that when an external magnetic field given by the vector potential ${\bf A}_{\rm em}$ is added,
the sum ${\bf A}_{\rm em}+{\bf A}_{\rm flux}$ is gauge invariant. For the gauge transformation
\begin{eqnarray}
{\bf A}_{\rm em} \rightarrow {\bf A}_{\rm em}+\nabla \phi,
\label{aem}
\end{eqnarray}
the total wave function transforms as
\begin{eqnarray}
e^{-i{1 \over 2}\sum_{j=1}^{N}\chi({\bf r}_j)}\Psi({\bf r}_1, \cdots, {\bf r}_N) 
\rightarrow e^{i\sum_{j=1}^{N}{{q\phi({\bf r}_j)} \over {\hbar c}}}
e^{-i{1 \over 2}\sum_{j=1}^{N}\chi({\bf r}_j)}\Psi({\bf r}_1, \cdots, {\bf r}_N),
\end{eqnarray}
thus, the phase $\chi$ transforms as
\begin{eqnarray}
\chi \rightarrow \chi-{{2q\phi} \over {\hbar c}};
\end{eqnarray}
which leads to the transformation of ${\bf A}_{\rm flux}$ as
\begin{eqnarray}
{\bf A}_{\rm flux} \rightarrow {\bf A}_{\rm flux}-\nabla \phi.
\label{aflux}
\end{eqnarray}
Combining Eqs.~(\ref{aem}) and (\ref{aflux}), it is deduced that the sum  ${\bf A}_{\rm em}+{\bf A}_{\rm flux}$ is gauge invariant; in other words, the phase ${1 \over 2}\chi$ is the {\em superpotential} of F. London\cite{London} that absorbs the arbitrariness in the gauge.

We can calculate the electric current using the energy functional in Eq.~(\ref{Drudef}) by substituting ${\bf A}_{\rm flux}+{\bf A}_{\rm em}$ for ${\bf A}_{\rm flux}$;
\begin{eqnarray}
&&E_{\rm tot}[{\bf A}_{\rm flux}
+{\bf A}_{\rm em}] \approx E_{\rm tot}[{\bf A}_{\rm flux}^{o}]
\nonumber
\\
&+{1 \over 2}& \int d^3 x \ d^3y 
\ ^{t}\!\left(\delta{\bf A}_{\rm flux}({\bf x}) +{\bf A}_{\rm em} ({\bf x}) \right){{\delta^2 E_{\rm tot}[{\bf A}_{\rm flux}^{o}]} \over {\delta {\bf A}_{\rm flux}({\bf x})
\delta {\bf A}_{\rm flux}({\bf y})}} \left(\delta{\bf A}_{\rm flux}({\bf y})+{\bf A}_{\rm em}({\bf y})\right),
\label{currentM}
\end{eqnarray}
where $\delta{\bf A}_{\rm flux}+{\bf A}_{\rm em}$ is regarded as a small perturbation. 
The expansion above usually does not exist if the gauge potential ${\bf A}_{\rm flux}$ is absent because the arbitrariness in the gauge of ${\bf A}_{\rm em}$ makes such an expansion meaningless; however, thanks to the presence of ${\bf A}_{\rm flux} $, the total energy is now a functional of the gauge invariant sum ${\bf A}_{\rm flux}+{\bf A}_{\rm em}$, thus, the above expansion does not depend on the choice of the gauge and is meaningful.
In Eq.~(\ref{currentM}), the term $\delta {\bf A}_{\rm flux}$ is added
to take into account the slight change of ${\bf A}_{\rm flux}$ from ${\bf A}_{\rm flux}^o$ due to the presence of the magnetic field $\nabla \times {\bf A}_{\rm em}$; thereby, the expansion becomes the one in terms of
$\delta{\bf A}_{\rm flux}+{\bf A}_{\rm em}$.

Using the functional Eq.~(\ref{currentM}),  the electric current is calculated as
\begin{eqnarray}
{\bf j}({\bf x})=-c \int d^3y 
{{\delta^2 E_{\rm tot}[{\bf A}_{\rm flux}^{o}]} \over {\delta {\bf A}_{\rm flux}({\bf x})
\delta {\bf A}_{\rm flux}({\bf y})}} \left(\delta{\bf A}_{\rm flux}({\bf y})+{\bf A}_{\rm em}({\bf y})\right).
\end{eqnarray}
This is gauge invariant since $\delta{\bf A}_{\rm flux}+{\bf A}_{\rm em}$ is gauge invariant.  It is known that the current given above leads to the Meissner effect, which is in accord with the fact mentioned in Section V that the state with the diamagnetic current becomes more stable when an external magnetic field is applied.

The Meissner effect means that the current is zero deep inside a large enough sample. 
Then, we have the relation, ${\bf A}_{\rm em}=-\delta {\bf A}_{\rm flux}=-{{c\hbar} \over {2q}}\nabla \delta \chi$, which leads to the flux quantization
\begin{eqnarray}
\Phi=\oint {\bf A}_{\rm em}\cdot d{\bf r}=-{{c\hbar} \over {2q}}\oint \delta \nabla \chi \cdot d{\bf r}
={{ch} \over {2q}}n,
\end{eqnarray}
where $n$ is an integer. In order to have a nonzero $n$, the conservation of the winding number must be lifted. Actually, when the quantized flux ${{ch} \over {2q}}$ occurs, the spin texture is the one produced by an isolated meron or antimeron located at the core region of the quantized flux.

Next let us consider the ac Josephson problem. The time variation of the phase of a wave function plays a role of the  chemical potential. When a Josephson junction is constructed with the present system, the time derivative of the 
phase difference $\Delta \chi$ across the junction is related to the chemical potential difference across the junction; namely, the voltage $V$ across the junction is related to $\Delta \dot{\chi}$ as
\begin{eqnarray}
{ 1 \over 2}\Delta \dot{\chi}=qV/\hbar.
\end{eqnarray}

The solution to the above equation is
\begin{eqnarray}
\Delta \chi={{2qV} \over {\hbar}}t +\mbox{const}.
\end{eqnarray}
Since $\chi$ and $\chi+2\pi$ are physically the same,
the above relation implies the current across the junction will oscillate with the frequency
$2qV/h$.  Overall, the flux quantization in the units ${{ch} \over {2q}}$ and the Josephson frequency ${{2qV} \over h}$ are both explainable in the present model.

The phase of the superconducting order parameter is currently believed to stem from the breakdown of the global $U(1)$ gauge invariance;\cite{Anderson} and it is also believed the particle number nonconserving formalism is necessary to take into account it.
However, the essential features of superconductivity are obtainable if a phase mode that parameterizes $U(1)/Z_2$ exists;\cite{Weinberg} thus, superconductivity without the breakdown of the global $U(1)$ gauge invariance is in principle possible. Indeed, the present theory shows such a phase, ${1 \over 2}\chi$ arises spontaneously without the breakdown of the global $U(1)$ gauge invaraince.

\section{Conclusion}
We have considered persistent current generation by small polarons embedded in a spin-density wave background in a two dimensional lattice. A spin current is generated by a spin Berry phase arising from a spiral spin texture produced by merons or antimerons. This state is currentless due to the exact cancellation of the currents for up- and down-spin electrons. However, when the polarons form side-by-side bipolarons, the motion of the system as a whole occurs, spontaneously; and the persistent current appears with the bipolaronic cores of merons and antimerons serving as {\em pumps} for it. The present model explains qualitatively various anomalous features of the cuprate superconductors. It also suggests the appearance of the phase variable or ``Nambu-Goldstone'' mode for the superconductivity without the breakdown of the global $U(1)$ gauge invariance.


\end{document}